# Growth Pattern of Magnetic Field Treated Bacteria

Samina Masood [a,*], Iram Saleem [b], Derek Smith [a] and Wei-Kan Chu [b]

[a] *Department of Physical and Applied Sciences, University of Houston-Clear Lake, Houston, TX, 77058, USA .*

[b] *Department of Physics and Texas Center for Superconductivity, University of Houston, Houston, TX, 77204, USA*

[*] Corresponding author email address: masood@uhcl.edu

**Abstract**. A study of the induced effect of different types of weak magnetic field exposure on bacterial growth is performed, comparing the relative changes after removal from the magnetic fields. This investigation is relevant to understand the effect of magnetic field exposure on human beings due to electronic devices. For this purpose, we use four species of common bacteria in reference to human health and safety including *Escherichia coli, Staphylococcus aureus, Staphylococcus epidermidis,* and *Pseudomonas aeruginosa.* The choice of these four bacteria also allows us to check for effects which rely upon the Gram-staining properties or shapes of bacterial species. These species were initially exposed to static, non-homogeneous and alternating weak magnetic fields, and then they were grown in incubators in the same environment at 37 °C simultaneously. Comparative measurements of optical density are then used to track the sustained impact on bacterial growth in the experimental samples. Bacteria was first grown in different weak magnetic fields on a plain glass surface both in liquid and solid media. Magnetic field treated bacteria were then transferred into similar test tubes to grow in an incubator concurrently. Bacterial cultures in liquid nutrient broth on plain glass proliferated faster in most species. Different magnetic fields affect the growth pattern of bacteria differently, depending on the bacterial strain. The weak magnetic field seems to decelerate the growth rate, even after the magnetic field is removed. With application of this study, we can potentially investigate the effect of weak field exposures on Eukaryotic cells and gene dynamics as well.

**Importance of this research**

Human body is commonly exposed to weak electromagnetic fields. Short term exposure of small fields may be ignorable but long term exposures may have significant effect. The ubiquitous nature of bacterial influence on the human bodies and the environment is well-accepted. We study the most commonly found bacteria in human body as residents or invaders, to understand the sustainable impact of weak field exposure. In this paper we specifically investigate, for the first time, that if the magnetic field effect is associated with the field or is sustained even after turning off the magnetic field. Different bacterial strains show different sustainable effect. It depends on the shape and gram-staining properties and will be applicable to improve the treatment of infectious diseases, cleaning environment and other medical applications. This result can also help in developing better methods to preserve food for a longer period of time.

**1. Introduction**

Bacterial adhesion and growth is affected by many factors including topography and material of the surface, associated ion flow, environmental conditions, and chemical features. Proliferation of the adherent bacteria and synthesis of the Exopolysaccharides (EPS) matrix is influenced in the presence of a weak magnetic field. It is therefore important for us to understand the interaction between biological systems and the magnetic fields which influence them [1, 2]. Bacterial cells are enclosed by a few nanometers thick cell membrane that is made up of ion channels which regulate ion flow through the membrane. The movement of these ions across a membrane forms the basis for survival and facilitates the growth of a cell. Living organisms have been exposed to electric and geomagnetic fields since the beginning of life on Earth, and numerous effects of magnetic fields have been reported on these organisms in the last few decades [3-5]. Despite this, there is not currently enough information about the dynamics involved in bacterial proliferation in the presence of a weak magnetic field.

Earth's magnetic field is weak and almost uniform, which can be ignored, whereas the current variety of weak magnetic fields generated by commonly used electrical appliances and devices are totally unavoidable. The role of bacterial growth in the formation of biofilms and the impact of weak magnetic field exposure has been of considerable interest in the food and healthcare industry, and is therefore a relevant topic for academic research. It was recently discovered that magnetic and electric fields of particular frequencies can have an influence on physiology or behavior of a cell [6-10]. Both stimulatory and inhibitory effects of weak fields on the growth rate has been observed during several investigations [11, 12]. Certain cells have magnetic structures which cause their enzymatic activities and RNA machinery to function differently in the presence of a magnetic field. This affects the growth rate, mutation frequency and other mechanisms. It has also been suggested that magnetic fields inactivate pathogenic microbes [13, 14]. A number of studies report inhibited growth of bacteria under the influence of strong electromagnetic fields [15-19]. However, a comprehensive study of the effect of weak magnetic fields on the growth and behavior of bacteria and a comparative study of this effect in relation to their gram-staining properties or their shape is still under investigation. Sustainable effects of weak magnetic fields have not been studied in detail either. In this paper, we present the results of an experimental investigation on how weak static, non-homogenous and alternating magnetic fields can affect the growth dynamics of various bacterial species. Several bacterial species were cultured in the same concentration of nutrient at the same temperature simultaneously. In addition, all test samples were always prepared out of the same mother cultures to minimize the changes due to extraneous environmental conditions.

Initially, all culture plates, including control and magnetic field-exposed cultures, were simultaneously grown at room temperature. Two Gram-negative (*Escherichia coli, Pseudomonas aeruginosa*) and two Gram-positive (*Staphylococcus epidermidis*, *Staphylococcus aureus*) species were exposed for three days in four different magnetic field conditions (described in detail in the materials and methods section) using liquid nutrient broth and agar as the growth media. Control and exposed samples of the four bacterial strains from both solid and liquid media were then regrown simultaneously in test tubes inside in incubator at 37

°C. Growth rates were discerned by optical density (OD) measurements at 650 nm for three days. Magnetic field conditions affect the growth rate of bacteria originally grown in liquid nutrient broth and agar environments differently.

This paper is organized such that the next section is devoted to the description of the experimental procedure, and contains information about the experimental setup and all materials used. All the results of this experiment are reported in section 3 with the discussion of our investigation in detail. In the last section, the main results of the paper are summarized and concluded.

**2. Materials and Methods**

*2.1 Bacteria and chemicals*

Bacterial cultures were obtained from Carolina Biological Supply Company, Burlington, North Carolina, USA. Four bacterial strains were used: Staphylococcus aureus (coagulase positive), MicroKwik Culture®, Pathogen, Vial (155554A), Staphylococcus epidermidis, MicroKwik Culture®, Vial (155556A) as Gram-positive bacteria, Pseudomonas aeruginosa, MicroKwik Culture®, Vial (155250A) and Escherichia coli, MicroKwik Culture® , Vial (155065A), as Gram-negative bacteria.

*2.2 Magnetic field setup*

We prepared four different magnetic field configurations in order to search for varying effects. The first of these was created by aligning an alternating series of twelve bar magnets in a line such that each pole is adjacent to the opposite poles as shown in Fig 1. Each individual bar magnet was approximately 0.75 inches wide and 6 inches long, building the total dimension of the rectangular arrangement to approximately 9 inches wide and 6 inches long. This magnetic field arrangement created a field strength according to position that stays in the range of |3G| to |5G|.

The second and simplest magnet configuration was made using eight individual small, circular and relatively strong magnets. These magnets had a diameter of approximately 1 and 7/16 inches. Each magnet was numbered individually, and kept separate for use and measurements. Because of the simplicity of fields generated by each individual round magnet, an approximate field strength can be measured without need for ranges of values. The approximate field measurements for each round magnet was about 75G at the center, and the magnets can be seen in Fig 2.

For the uniform magnetic field setup, we used a series of eight 200 turn coils laid in order to achieve a vertical oriented cylinder configuration. One terminal of a direct current power supply was plugged into the bottom coil and the other was connected to the top coil. This creates what is effectively a large solenoid, allowing for a nearly uniform field to develop within the interior region of the coils. Using a magnetic field sensor, we tuned the DC voltage to result in a magnetic field of approximately 5 Gauss in the center of the solenoid. This uniform magnetic field setup can be seen in Fig 3.

The oscillating magnetic field was set up in a manner similar to the configuration used for the uniform field. The difference from the previous setup is that this one was connected to a programmable power supply, which allowed for the magnetic field to vary with time. The system was set to change field strength once every minute by alternating between two voltage settings every 60 seconds. The two settings were tuned to allow for a near-constant field of strengths approximately 0.5 Gauss and 5 Gauss, with 0.5 Gauss being detected when the system is on its lowest setting. This slowly varying, oscillatory magnetic field setup can be seen in Fig 4. We grew the control bacteria in the same room at the same time as the experimental samples in order to ensure all extraneous environmental conditions are similar for each.

*2.3 Experimental Procedure*

Experimental samples were prepared using pipettes and sterilized plates. Half of the plates were filled with agar, while the other half remained unfilled. Prior to inoculation the unfilled plates had a sterilized, flat circular glass coverslip placed into the center of the plate. Each set of plates was then inoculated with bacteria pipetted from the mother culture, including the plates not containing agar. After inoculation, sterilized glass cover slips were placed on the site of inoculation on the agar plates. The two types of media were chosen to illustrate if the bacterial growth depends on the form of the growth media. After each plate was carefully labeled, the experimental samples were then placed at their respective positions in the previously described four magnetic field environments, and the control samples were placed separately. After three days of growing at room temperature in their respective environments, the plates were then removed in order to be processed.

Each agar plate was scraped for bacteria by carefully targeting one individual colony, which was then inoculated into the nutrient broth in sterile tubes. When the measurement process was ready to begin, each sample was vortexed and 50µl was taken from each test tube and inoculated into three fresh test tubes with 5ml of broth. After preparation was complete, the numerous low concentration experimental samples were then placed into a shaking incubator set to 37 °C. The samples were then periodically removed from the incubator for a short period of time in order to take absorption readings with an optical spectroscope. After each test tube was individually measured, the entire set of bacteria was then placed back into the incubator until the time of the next scheduled measurement, which was approximately once every two hours. This process of taking periodic measurements continued for 24-36 hours, depending on the bacterial strain.

The purpose of the present study is to investigate the sustained effect of different weak magnetic field conditions upon the growth rate of bacteria outside the magnetic field. The effect of specific magnetic fields on the growth pattern of the various strains of bacteria was demonstrated by optical density measurements. The device used to take optical measurements of the bacteria was a Spectronic 20D+. 13 x 100 mm Pyrex test tubes (with caps) were employed for this experiment in order to fit into the spectrometer. Optical properties of the nutrient broth mixtures change as the bacteria grows in a test tube, and the pure broth had a minimum in absorbance at 650 nanometers. This allowed us to track the increase of absorbance as an indicator of bacterial growth. The spectrometer was set to take measurements at 650 nm, and calibrated using "blank" test tubes containing only pure broth.

In this paper we primarily focused on the impact of weak magnetic fields on the growth rate of bacteria until reaching the stationary phase. Thus our main emphasis is the log phase of the bacterial growth curves.

## 2. Results and discussion

We plot the growth curves for each bacterial species which was originally grown in different types of magnetic fields, both in nutrient broth and on agar with a glass coverslip, under the same environmental conditions. These results are also compared with the results from their corresponding control cultures, which were never exposed to a magnetic field. All samples chosen for comparison were grown concurrently in an incubator set to 37 °C.

To investigate the sustained effect of a magnetic field on bacterial growth, we plot the optical density of bacteria as a function of time for almost 36 hours to clearly distinguish between the effects of different types of fields on different species.

Fig. 5 shows a comparison between the influence of different types of magnetic fields on the bacterial growth of *Escherichia coli* (*E. coli*), both in nutrient broth (liquid media) and agar (solid media) in a comparable concentration of nutrients. Magnetic field-exposed samples have generally slower growth rates. The bar magnets show a small effect on the growth rate, comparable to the control samples. On the other hand, the oscillating magnetic field has a more significant effect on the growth rate, as minimal growth is seen in the bacteria treated by the oscillating field. It is also noticed that the growth patterns are not significantly affected, as the general trend of a bacterial growth curve is followed by each set. However, the difference from the control grows with time, indicating that the inhibiting effect of a magnetic field maintains over several generations of bacteria.

In Fig 6, the data sets from both agar and nutrient broth samples are plotted together to compare the effect of growth medium in each type of magnetic field individually. A clear difference in bacterial growth between nutrient broth and agar can be seen, as samples grown in agar reach the saturation level significantly faster. The difference in the growth curves represents the effect of the nature of the medium and the type and strength of the magnetic field on the rate of bacterial cell division. Initially the bar magnetic field samples grow with much less variance from the control samples, but after about 6 hours the exposed sample curves begin to split off from the control. Bacteria which is grown either in the relatively strong fields of the round magnets or in the oscillating field seem to take a relatively longer amount of time to reach the stationary phase. This effect is expected to be coincidental, as the fields are not similar in behavior. The round magnets have relatively strong fields, which will have a more pronounced inhibiting effect when compared to similar but weaker fields such as that of the bar magnets.

The lag phase is experienced in a large proportion of the oscillating magnetic field samples, which can be explained as the bacteria taking a longer time to adapt to the oscillating magnetic field environment. The bacteria in the oscillating magnetic field takes more time to acclimate to its new environment since the magnetic field changes regularly. Because of this, oscillating field samples take longer to reach significant levels of growth. As expected, the growth rate appears to be slower for the agar samples as compared to the bacteria grown in liquid nutrient broth. Justification of this behavior is related to the mobility of nutrients in the liquid broth as compared to the jelly-like agar. This mobility keeps increasing the difference in growth rate with time. It is also clearly seen that the difference between the control and the magnet-exposed bacterial growth is more pronounced as compared to the difference in a liquid and solid medium.

Similar kinds of results are obtained for the Gram-positive bacteria *Staphylococcus aureus* (*S. aureus*) in the sense that magnetic field exposure suppresses the growth overall. However, *S. aureus* seems to enter into the stationary phase more quickly than the Gram-negative *E. coli*. Fig 7 shows that for *S. aureus*, the gap among the bacterial growth curves in all the four different magnetic field configurations is more distinguishable than *E. coli*. The oscillating field inhibits the growth to a much stronger extent in comparison with the control. As can be seen, the gap between the two curves is large. This is probably because the bacteria grows to its maximum level faster, and thus the impact is significant in a short time and the samples have a larger amount of variance to them compared to other samples.

In Fig 7b, the growth curve for uniform and bar magnetic fields are almost overlapping in the beginning. But after 8 hours, the behavior is distinguishable and the uniform field has a stronger inhibiting influence on the growth. In general, the growth curves for magnetic field treated bacteria separate after approximately 4 hours. The figures also demonstrate that *S, aureus* distinguishes between different magnetic field influences more prominently as compared to *E. coli*.

Fig 8 shows the difference of growth rate between solid and liquid media in the same magnetic field in comparison with the control. It is interesting to notice that *S. aureus* does not show much difference between the behavior of agar and nutrient broth in all types of fields. However, nutrient broth growth is faster than agar.

Fig 9 shows a similar set of graphs for *Staphylococcus epidermidis* (*S. epidermidis*). It is interesting to note in this case how oscillating magnetic field exposed bacteria do not always exhibit the least growth, as the round and oscillating magnetic field growth curves are comparable in value. The inhibited growth rate in the oscillating field samples may be due to the alternating field effect which does not give bacteria enough time to adapt to its changing environmental conditions, and not due to the strength of the magnetic field. Conversely, the round magnetic field samples are affected simply due to the higher strength of magnetic fields. Coincidentally, the data from the round magnet-exposed samples is similar to the data obtained from the oscillating field samples, except for some slight variance. The data therefore strongly agrees with the possibility of a delayed effect on growth rate.

Fig 10 shows relatively more suppression in growth due to exposure to the magnetic fields in agar samples, but the separation between nutrient broth and agar growth curves is still small compared to *E. coli*. This result is comparable to *S. aureus*. This therefore indicates that the type of growth media may play an important role in low magnetic field environments.

For *Pseudomonas aeruginosa* (*P. aeruginosa*), we studied bacterial growth in liquid nutrient broth on a glass coverslip (Fig 11). It can be seen that *P. aeruginosa* does not get influenced much by small changes in the field. The similarity of the uniform and bar magnet field data sets, as well as the small difference from the control sample growth curve, indicate a small change for static and uniform fields. Moreover, *P. aeruginosa* takes a longer time to start the process of cell division, experiencing a long lag phase, while also staying in the exponential phase for a longer time.

Lastly, we examine the data for the oscillating magnetic field environment. The curve for each of the experimental sample sets are very similar except for the oscillating growth curve, but each set can be seen to deviate from the control after a point. The experimental samples continued to grow up until the end of the measuring process, but the bacteria exposed to an oscillating field reached its stationary phase faster than the rest. As the largest difference between control and the magnetic field treated samples is seen here, the 0-5 Gauss oscillating field is a good candidate for further *Pseudomonas aeruginosa* investigations.

Fig.12, gives a comparison of the suppression in growth of different bacterial species in solid (red) and liquid (blue) media previously exposed to different magnetic field environments. The growth of the control samples, where each bacteria is grown outside the field, is normalized to one. It is a common feature of these curves that the solid medium (agar) has a lower maximum value as compared to the corresponding liquid medium growth. The saturation values for the bacterial species shows that it is not only the strength and type of the magnetic field which matters in the growth rate, as different bacterial species are also affected differently according to the chosen growth media. The response of each bacterial species to the change in environmental conditions, such as magnetic field, may be potentially related to the differences in motility of individual species [20], in addition to their shapes and Gram staining properties.

## 4. Conclusion

Our study of the growth rates of different magnetic field-exposed bacterial strains shows that the magnetic field effect is retained by bacteria after removal from the magnetic field environments. The sustained inhibiting effect of previous exposure to magnetic fields was observed when the bacteria were grown under normal lab conditions outside of the experimental magnetic fields. The growth rate for a culture subjected to alternating magnetic fields is significantly reduced in comparison with that of the control cultures and other weak field-influenced bacteria. This result is consistent in all four bacterial strains. Overall, magnetic fields inhibit the growth of these bacterial species. However, the relatively strong round magnets have a pronounced effect.

Bacteria grown in agar over magnets show slowed growth compared to the samples grown in nutrient broth. We believe this general trend is due to the fluidity of nutrient broth, which increases the availability of nutrients as compared to agar medium. However, the variation in results may subject to the motility of the individual bacterial species This effect is more distinguishable in *E. coli*, which may be due to their adhesive properties [21] and swimming ability as compared to the different modes of motility of the other bacterial species. The separation in the growth curves of different magnetic fields is more pronounced in *S. aureus* as compared to other bacterial species. *P. aeruginosa* shows minimum influence on the growth curves for different magnetic field conditions, with the exception of the periodically changing magnetic field in which it reaches the saturation level much faster than the rest. This study can potentially be furthered to investigate the nature of effects on Eukaryotic cells, or even DNA dynamics.

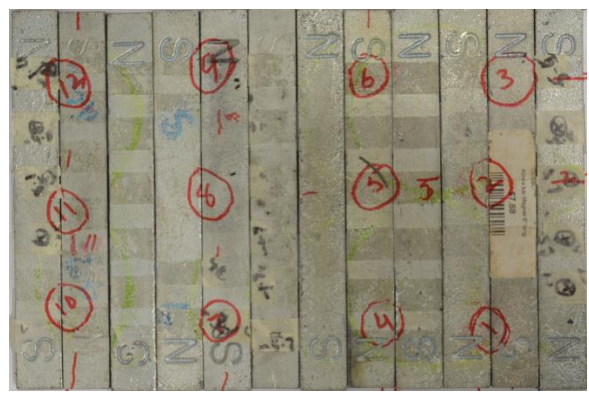

Fig 1: *Bar magnetic field arrangement.*

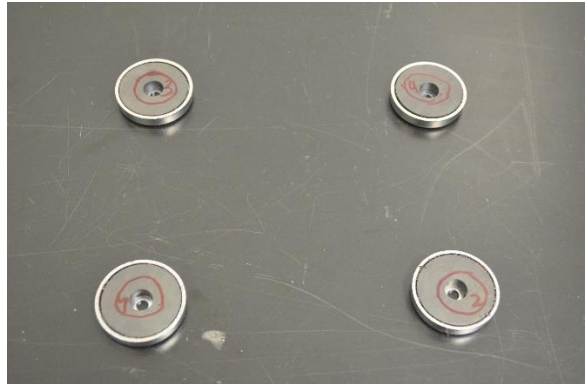

Fig 2: *Round magnets.*

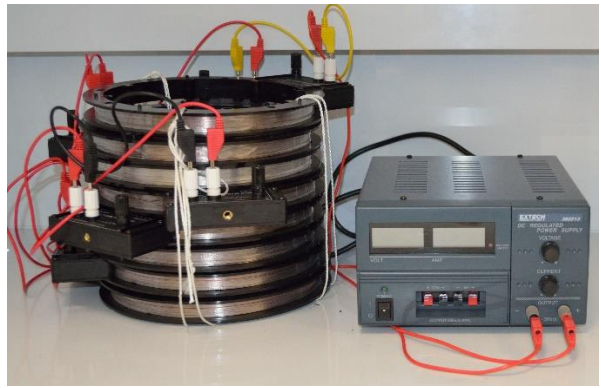

Fig 3: *Uniform field configuration.*

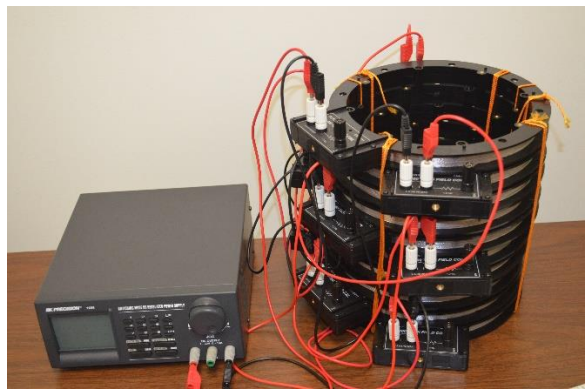

Fig 4: *Oscillating field configuration.*

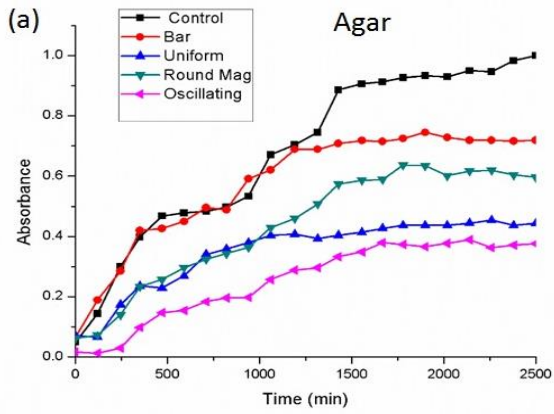
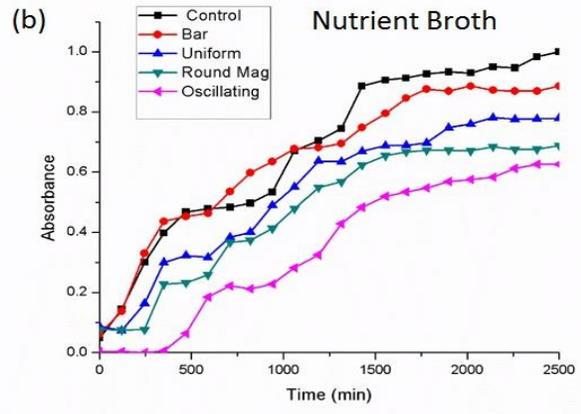

Fig 5: *Escherichia coli growth curves for (a) agar and (b) nutrient broth samples.*

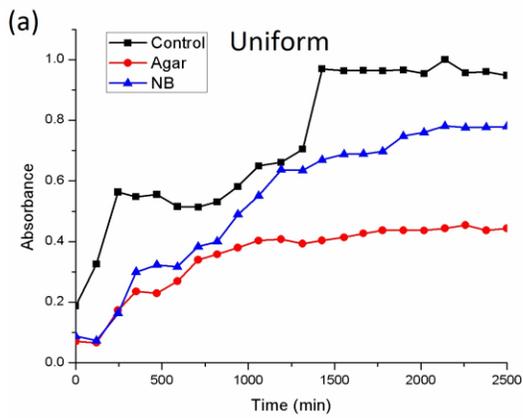
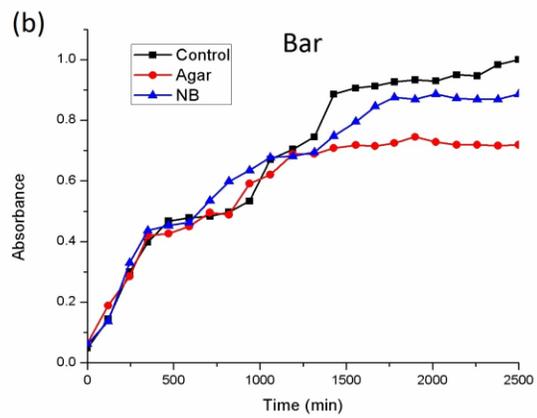
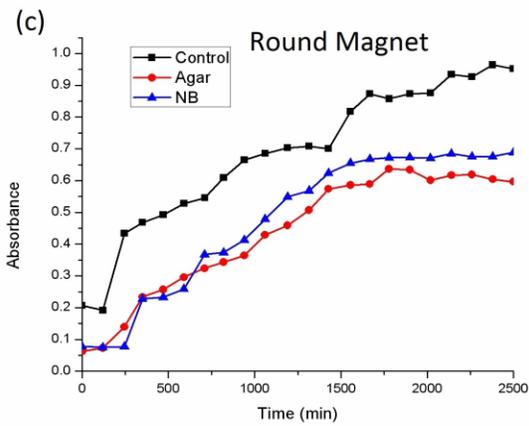
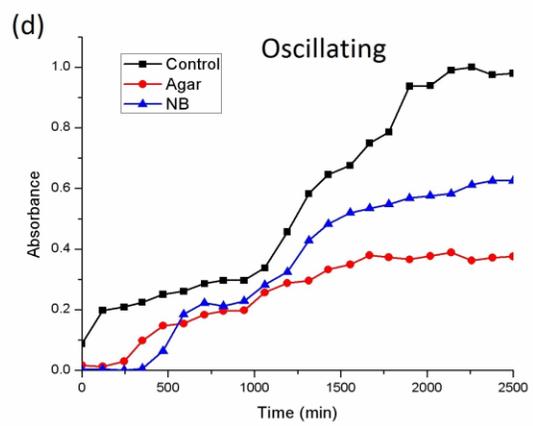

Fig 6: *Separated Escherichia coli growth curves for agar and nutrient broth samples in (a) uniform, (b) bar, (c) round magnet and (d) oscillating magnetic field setups.*

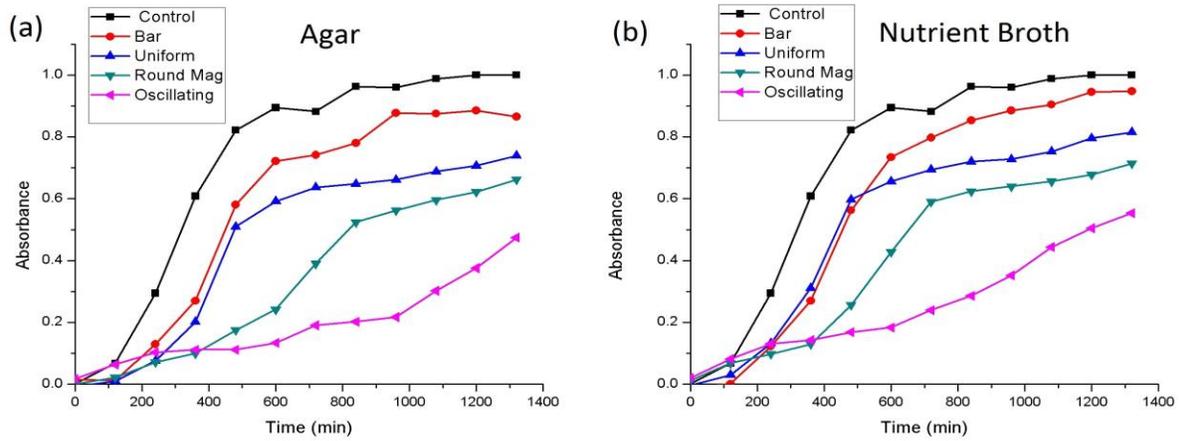

Fig 7: *Growth curve of magnetic field treated* Staphylococcus aureus*, in (a) agar and (b) nutrient broth.*

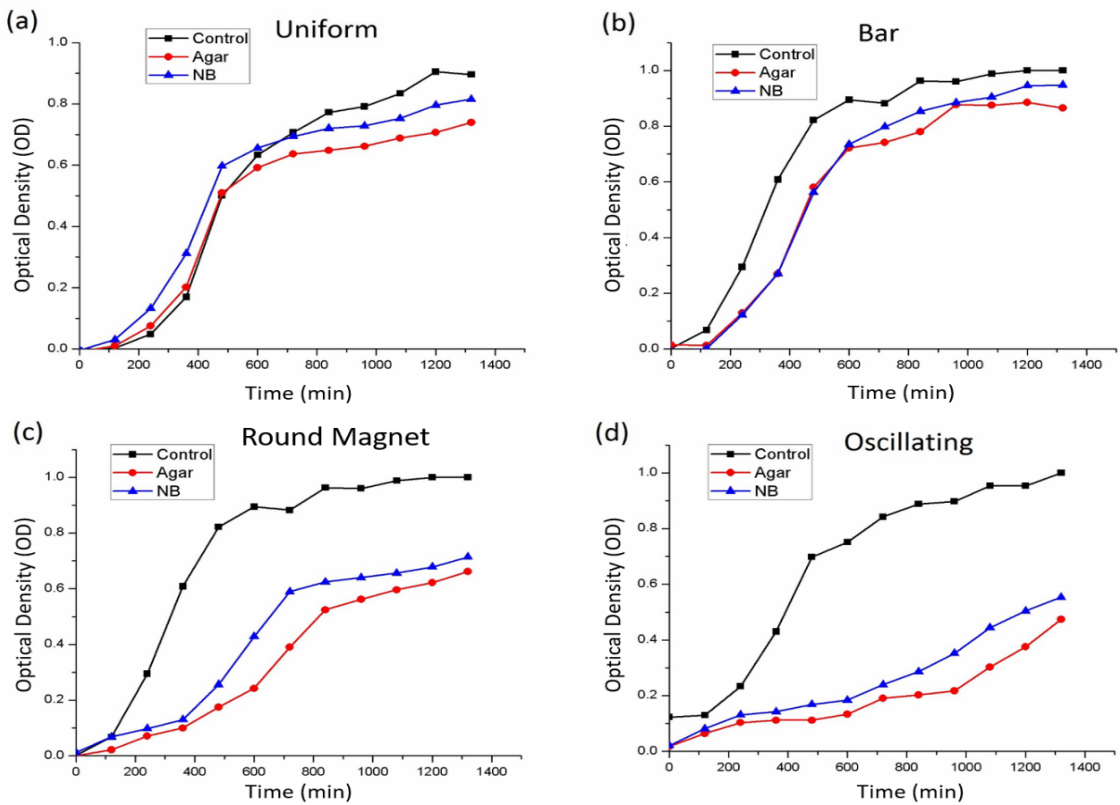

Fig 8: *Comparing* Staphylococcus aureus *growth curves for agar and nutrient broth samples in (a) uniform, (b) bar, (c) round magnet and (d) oscillating magnetic fields.*

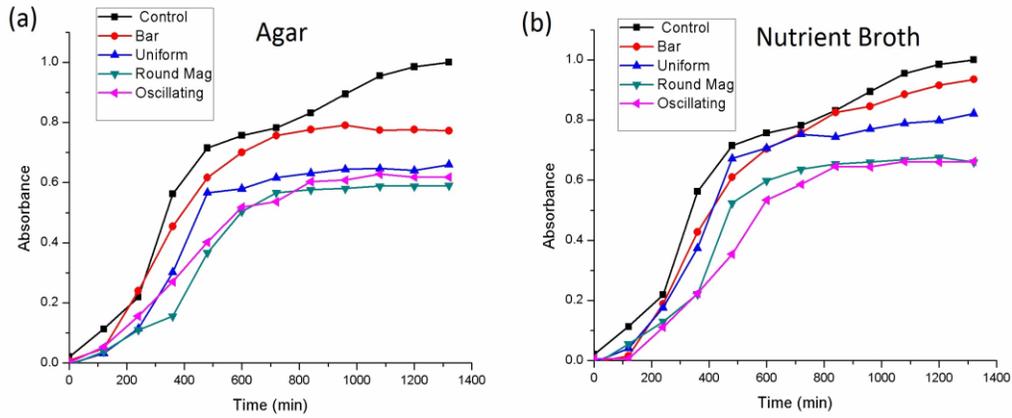

Fig 9: *Staphylococcus epidermidis growth curves for (a) agar and (b) nutrient broth samples.*

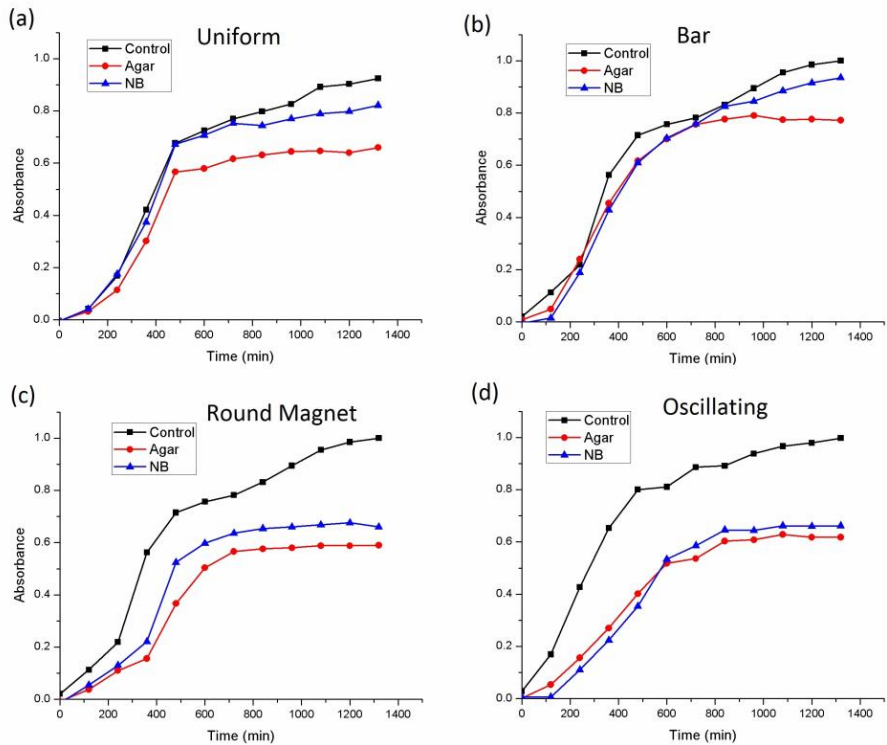

Fig 10: *Comparing Staphylococcus epidermidis growth curves for agar and nutrient broth samples in (a) uniform, (b) bar, (c) round magnet and (d) oscillating magnetic fields.*

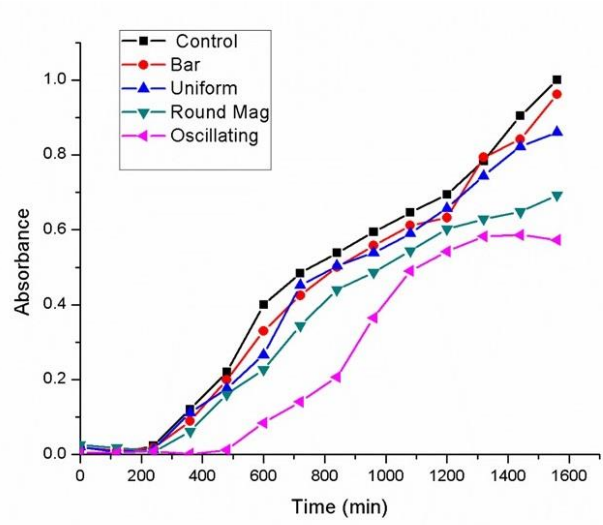

Fig 11: Pseudomonas aeruginosa growth curves for nutrient broth samples.

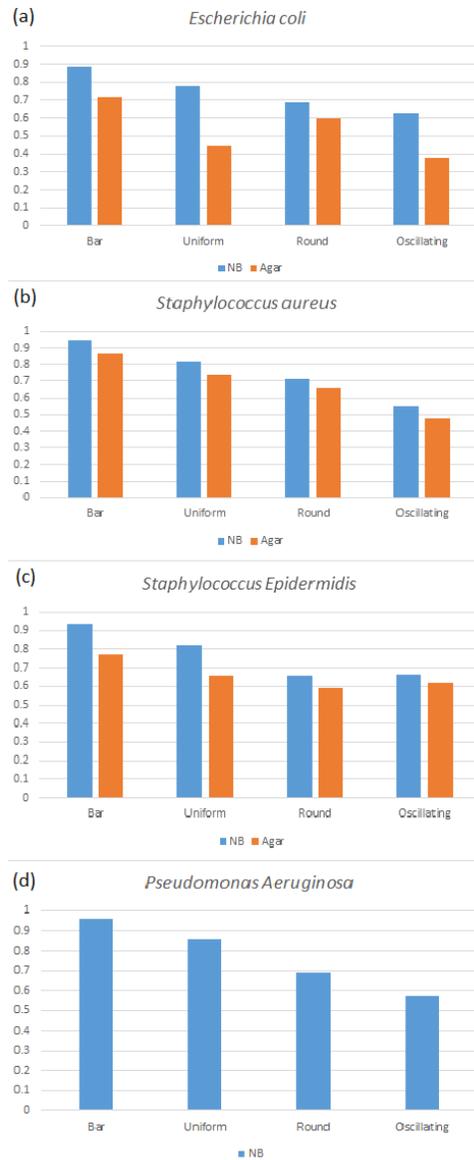

Fig 12: *Maximum optical density value during growth period of (a) Escherichia coli, (b) Staphylococcus aureus, (c) Staphylococcus epidermidis and (d) Pseudomonas aeruginosa in bar, uniform, round and oscillating magnetic fields for nutrient broth (in blue) and agar (in red). Pseudomonas aeruginosa was studied in liquid medium alone. All samples normalized such that the controls would have a value of one.*